\documentstyle[12pt,fleqn]{article}   
\newcommand{\preprint}[1]{\hfill{\sl preprint - #1}\par\bigskip\par\rm}
\renewcommand{\title}[1]{\begin{center}\Large\bf #1
 \end{center}\rm\par\bigskip}
\renewcommand{\author}[1]{\begin{center}\Large #1\end{center}}
\newcommand{\address}[1]{\begin{center}\large #1 \end{center}}

\def\dinfn{\smallskip Dipartimento di Fisica, Universit\`a di Trento\\ 
                           and Istituto Nazionale di Fisica Nucleare,\\
                                   Gruppo Collegato di Trento, Italia}

\def\Idinfn{\address{\dinfn}}
\def\references{\begin{thebibliography}{10}}
\def\endreferences{\end{thebibliography}\end{document}}

\renewcommand{\date}[1]{\par\bigskip\par\sl\hfill #1\par\medskip\par\rm}
\newcommand{\email}[1]{e-mail: \sl #1@science.unitn.it\rm}
\newcommand{\femail}[1]{\footnote{\email{#1}}}
\newcommand{\pacs}[1]{\smallskip\noindent{\sl PACS number(s):
                       \hspace{0.3cm}#1}\par\bigskip\rm}
\def\babs{\hrule\par\begin{description}\item{Abstract: }\it} 
\def\eabs{\par\end{description}\hrule\par\medskip\rm}
 
\renewcommand{\vec}[1]{{\bf #1}}       %%%  vectors in bold
\def\hs{\qquad\qquad}         %%%  horizontal space
\def\nn{\nonumber}            %%%  no number for eqnarray
\def\beq{\begin{eqnarray}}    %%%  begequation/eqnarray
\def\eeq{\end{eqnarray}}      %%%  endequation/eqnarray
\def\R{\mbox{$I\!\!R$}}                 %%% real numbers
\def\ii{\infty}                         %%% infinit
\def\al{\alpha}
\def\be{\beta}
\def\ga{\gamma}
\def\de{\delta}
\def\ep{\varepsilon}

\def\si{\sigma}
\def\om{\omega}

\def\La{\Lambda}

\begin{document}
%\tableofcontents       %%%%%%   index of section

\preprint{UTF 352 \\ hep-th/9507139} 
\title{Thermal Wightman Functions and Renormalized Stress Tensors in the 
  Rindler Wedge}

\author{Valter Moretti \femail{moretti}, Luciano Vanzo \femail{vanzo}}

\Idinfn

\date{July - 1995}

\begin{abstract}
The Wightman functions in the Rindler portion of Minkowski
space-time are presented for any value of the temperature and for 
massless spin fields up to $s=1$ and the renormalized stress tensor 
relative to Minkowski vacuum is discussed. A gauge ambiguity in the 
vector case is pointed out.
\end{abstract}


%\bigskip

\section{Wightman Functions and Stress Tensors}

The Rindler regions can be defined with respect to
any spacelike two-plane ${\cal P}$ in Minkowsi space-time. We may 
choose rectangular coordinates $(x,y,z,t)$ such that the plane is the set  
$x=t=0$. The Rindler wedge we shall consider, denoted $W_R$, will then 
be defined by the inequality $x>|t|$. A global parametrization of 
$W_R$ is obtained by setting $x=\xi\cosh\tau$, $t=\xi\sinh\tau$, 
for $\xi>0$, so 
that $x^2-t^2=\xi^2$. Thus any line $\xi=\xi_0$, $y=y_0$, $z=z_0$
will be the trajectory of a uniformly accelerated particle, with 
proper acceleration $a=\xi_0^{-1}$ and proper time $s=a\tau$ along the
trajectory. The Minkowski metric will take 
the form $ds^2=-\xi^2d\tau^2+d\xi^2+dx_t^2$, with $\xi>0$ and 
$x_t=(y,z)$ standing for the transverse coordinates. The metric admits 
the timelike Killing field $K=\partial_{\tau}$ generating the 
isometry $\tau\rightarrow\tau+\tau_0$.  
The hypersurface $\xi=0$ is an event horizon which bifurcates in the 
transverse two-plane ${\cal P}$.  

We shall find the thermal Wightman functions in 
the Rindler region $W_R$ (the left region $W_L$ is then covered by 
reflecting through the wedge, namely by sending 
$(t,x,x_t)\rightarrow(-t,-x,x_t)$). Hence it is understood that
fields quantization in this region is defined by taking the 
Fock representation over a vacuum $|F>$ which is invariant under 
translations in $\tau$ (it is customary to call $|F>$ the Fulling 
vacuum\cite{full73-7-2850,full77-10-917,unru76-14-870,boul75-11-1404}  
(for an alternative description of accelerated systems, see 
\cite{sanc86-34-1056}). 

The vacuum Wightman functions for a general field 
$\phi_A(x)$ are then defined as the expectation values 
\beq
W^+_{AB}(x,x^{'})=<F|\phi_A(x)\phi_B(x^{'})|F>, \hs
W^-_{AB}(x,x^{'})=<F|\phi_B(x^{'})\phi_A(x)|F>
\eeq
The definition will be the same for other states as well, in 
particular for the Minkowski vacuum $|M>$. The situation will be
rather different for a thermal equilibrium state, since then 
there is no obvious way to compute the expectation values. This is because 
the partition function for a quantum field is divergent in the infinite 
volume of the Rindler region. The thermal Wightman functions will then 
be defined as the periodic or anti-periodic solution of the field 
equations having the analyticy properties which are demanded by the 
KMS condition\cite{haag67-5-215,full87-152-135}. An independent 
check will be then to recover the 
vacuum expectation values in the limit $\be\rightarrow\ii$ of zero 
temperature. For future reference we define the quantity $\al$ by
\beq
\cosh\al=\frac{\xi^2+\xi^{'2}+|x_t-x^{'}_t|^2}{2\xi\xi^{'}}\nn
\eeq
The Weyl and electromagnetic fields will be defined with respect to 
the natural orthonormal vierbein
\beq
e_{(0)}^a=\xi^{-1}\de_0^a, \hs e_{(i)}^a=\de_i^a, \hs i=1,2,3\nn
\eeq
where $a,b,c,..$ denotes coordinate indices and $i,j,k,..$ 
anholonomic, or vierbein indices.
The thermal Wightman functions at inverse temperature $\be$ for a 
massless field with elicity $s>0$ will be denoted by 
$W^{(s)\pm}(\be|x,x^{'})$ and simply by $W^{\pm}(\be|x,x^{'})$ in the
spin zero case. We give them first and then we discuss how they were 
obtained. They are given as follows:\\
a) the scalar $s=0$ field
\beq
W^+(\be|x,x^{'})=\frac{1}{4\pi\be\xi\xi^{'}\sinh\al}
\left[\frac{\sinh\frac{2\pi}{\be}\al}{\cosh\frac{2\pi}{\be}\al
-\cosh\frac{2\pi}{\be}(\tau-\tau^{'}-i\ep)}\right]
\label{scal}
\eeq
and $W^{-}(\be|x,x^{'})=[W^{+}(\be|x,x^{'})]^*$. These are 
manifestly periodic in imaginary time with period $\be$. To our 
knowledge, this result was first obtained by J.S.Dowker
\cite{dowk77-10-115,dowk78-18-1856}.
%\begin{enumerate}
 The zero temperature limit is
\beq
W^+(x,x^{'})=<F|\phi(x)\phi(x^{'})|F>=-\frac{1}{4\pi^2}\frac{\al}{\xi
\xi^{'}\sinh\al}\frac{1}{(\tau-\tau^{'}-i\ep)^2-\al^2}
\eeq
and $W^-(x,x^{'})=<F|\phi(x^{'})\phi(x)|F>=[W^+(x,x^{'})]^*$. Note 
that these functions are vacuum expectation values in the Fulling 
state. 
 The thermal function can be obtained by the sum over images
\beq
W^{\pm}(\be|x,x^{'})=-\frac{1}{4\pi^2}\frac{\al}{\xi\xi^{'}\sinh\al}
\sum_{n=-\ii}^{\ii}\frac{1}{(\tau-\tau^{'}\mp i\ep-in\be)^2-\al^2}
\label{imag}
\eeq
 The value $\be=2\pi$ is distinguished by the property
\begin{eqnarray}
W^+(2\pi|x,x^{'})&=&\frac{1}{4\pi^2}\frac{1}{\xi^2+\xi^{'2}+
|x_t-x^{'}_t|^2-2\xi\xi^{'}\cosh(\tau-\tau^{'}-i\ep)} \nn \\ 
&=&-\frac{1}{4\pi^2}\frac{1}{(t-t^{'}-i\ep)^2-(x-x^{'})^2-|x_t-x^{'}_t|^2}
\end{eqnarray}
which is just the Wightman function which characterizes the Minkowski 
vacuum state. This means that this vacuum is a KMS state with respect to 
$\tau$-translation
\cite{haag84-94-219,full77-10-917,unru76-14-870,isra76-57-107,dowk77-10-115,
davi75-8-609}, with inverse temperature $\be=2\pi$.  
The thermal stress tensor relative to the Minkowski vacuum is
\cite{brow86-33-2840,brow85-31-2514,dowk87-36-3742}
\beq
T^{ab}=\frac{1}{1440\pi^2\xi^4}\left[\left(\frac{2\pi}{\be}\right)^4-1\right]
[4v^av^b+g^{ab}]
\label{tt}
\eeq 
where $v^a=K^a/\sqrt{-K^2}$, $K=\partial_{\tau}$ being the Killing 
vector field of the Rindler region. The zero temperature stress tensor 
reduces to the one calculated in Ref.\cite{cand77-254-79}.\\
b) Weyl $s=1/2$ fermions.\\
There are two irreducible representations 
of the Dirac algebra. Denoting by $\vec{\si}$ the Pauli matrices, 
these are given by $\si^i=(e,\vec{\si})$ and 
$\tilde{\si}^i=(e,-\vec{\si})$, where $e$ is the unit. In the tilde 
representation we find
\beq
W^{(1/2)+}(\be|x,x^{'})=i\tilde\si^a\tilde\nabla_aF^+_{\be}(x,x^{'})
\nn
\eeq
\beq
W^{(1/2)-}(\be|x,x^{'})=-i\tilde\si^a\tilde\nabla_aF^-_{\be}(x,x^{'})
\nn
\eeq
where
\begin{eqnarray}
F^+_{\be}(x,x^{'})&=&\frac{e}{4\pi\be\xi\xi^{'}\sinh\frac{\al}{2}}
\left[\frac{\sinh\left(\frac{\pi}{\be}\al\right)\cosh\left[\frac{\pi}{\be}
(\tau-\tau^{'})\right]}{\cosh\left(\frac{2\pi}{\be}\al\right)-
\cosh\left[\frac{2\pi}{\be}(\tau-\tau^{'}-i\ep)\right]}\right]\\ \nn
&+&\frac{\si_1}{4\pi\be\xi\xi^{'}\cosh\frac{\al}{2}}
\left[\frac{\cosh\left(\frac{\pi}{\be}\al\right)\sinh\left[\frac{\pi}{\be}
(\tau-\tau^{'})\right]}{\cosh\left(\frac{2\pi}{\be}\al\right)-
\cosh\left[\frac{2\pi}{\be}(\tau-\tau^{'}-i\ep)\right]}\right]
\end{eqnarray}
These are manifestly anti-periodic in imaginary time with period 
$\be$, in accord with the KMS condition.
%\begin{enumerate}
 The zero temperature limit is
\beq
W^{(1/2)+}(x,x^{'})=<F|\psi(x)\psi^{\dag}(x^{'})|F>=
i\tilde\si^a\tilde\nabla_aF^+(x,x^{'})\nn
\eeq
\beq
W^{(1/2)-}(x,x^{'})=-<F|\psi^{\dag}(x^{'})\psi(x)|F>=
-i\tilde\si^a\tilde\nabla_aF^-(x,x^{'})\nn
\eeq
where 
\begin{eqnarray}
F^+(x,x^{'})&=&-\frac{e}{8\pi^2\xi\xi^{'}\sinh\frac{\al}{2}}
\left[\frac{\al}{(\tau-\tau^{'}-i\ep)^2-\al^2}\right]- \nn \\ 
&-&\frac{\si_1}{8\pi^2\xi\xi^{'}\cosh\frac{\al}{2}}
\left[\frac{\tau-\tau^{'}}{(\tau-\tau^{'}-i\ep)^2-\al^2}\right]
\end{eqnarray}
and $F^-=(F^{+})^*$. Note that these functions are 
vacuum expectation values in the Fulling state $|F>$.  
The sum over images with alternating signs 
gives again the above thermal functions. 
The special value $\be=2\pi$ is distinguished, since then
\beq
W^{(1/2)\pm}(2\pi|x,x^{'})=\pm i\tilde\si^a\nabla_aW^{\pm}(x,x^{'})\nn
\eeq
where $W^{\pm}(x,x^{'})$ are the Wightman function for the massless scalar 
field. These are just the Wightman functions for neutrinos 
in the Minkoswki fermion vacuum, relative to a boosted tetrad.
 The stress tensor relative to the Minkowski vacuum 
has the perfect fluid form\cite{frol87-35-3779,dowk87-36-3742}
\beq
T^{ab}=\frac{1}{11520\pi^2\xi^4}\left[7\left(\frac{2\pi}{\be}
\right)^4+10\left(\frac{2\pi}{\be}\right)^2-17\right][4v^av^b+g^{ab}] 
\eeq
The zero temperature tensor was calculated in 
Ref.\cite{cand78-362-251} (see also Ref.\cite{troo79-159-442}).\\
c) The electromagnetic $s=1$ field.\\ 
 The tetrad components of the 
Wightman functions $W^{(1)\pm}_{ij^{'}}(\be|x,x^{'})$ will be given in 
the Feynman gauge $\nabla_aA^a=0$, where a prime over the indices 
means that the function is a bivector at $x$ and $x^{'}$ respectively. 
They are given by the equations 
\beq
W^{(1)+}_{00^{'}}(\be|x,x^{'})=\frac{-1}{4\pi\be\xi\xi^{'}\sinh\al}
\frac{\cosh\left(\frac{2\pi}{\be}
(\tau-\tau^{'})\right)\sinh\al+\sinh\left(\frac{2\pi}
{\be}-1\right)\al}{\cosh\left(\frac{2\pi}{\be}\al\right)-
\cosh\left(\frac{2\pi}{\be}(\tau-\tau^{'}-i\ep)\right)}
\eeq
\beq
W^{(1)+}_{11^{'}}(\be|x,x^{'})=-W^{(1)+}_{00^{'}}(\be|x,x^{'})
\eeq
\beq
W^{(1)+}_{10^{'}}(\be|x,x^{'})=
\frac{-1}{4\pi\be\xi\xi^{'}}\frac{\sinh\left(\frac{2\pi}{\be}
(\tau-\tau^{'})\right)}{\cosh\left(\frac{2\pi}{\be}\al\right)-
\cosh\left(\frac{2\pi}{\be}(\tau-\tau^{'}-i\ep)\right)}
\eeq
\beq
W^{(1)+}_{01^{'}}(\be|x,x^{'})=-W^{(1)+}_{10^{'}}(\be|x,x^{'})
\eeq
\beq
W^{(1)+}_{22^{'}}(\be|x,x^{'})=W^{(1)+}_{33^{'}}(\be|x,x^{'})
=W^+(\be|x,x^{'})
\eeq
where $W^+(\be|x,x^{'})$ is the scalar Wightman function, 
Eq.~(\ref{scal}). The periodicity in imaginary time is again evident.
In Ref.\cite{dowk87-36-3742} the Green functions for any spin 
around a cosmic string have been given in the $(j,0)$ representation 
of the Lorentz group. For $s=1$ elicity fields these are Green 
functions for the fields $\vec{E}\pm i\vec{H}$, and subtle questions 
of gauge invariance were consequently avoided. 
%\begin{enumerate}
 The zero temperature limit is
\beq
W^{(1)+}_{00^{'}}(x,x^{'})=-W^{(1)+}_{11^{'}}(x,x^{'})=
\frac{-1}{4\pi^2\xi\xi^{'}}\frac{\al\coth\al}{\al^2-(\tau-\tau^{'}-i\ep)^2}
\label{zero}
\eeq
\beq
W^{(1)+}_{10^{'}}(x,x^{'})=-W^{(1)+}_{01^{'}}(x,x^{'})=
\frac{-1}{4\pi^2\xi\xi^{'}}
\frac{\tau-\tau^{'}}{\al^2-(\tau-\tau^{'}-i\ep)^2}
\label{nozero}
\eeq
\beq
W^{(1)+}_{22^{'}}(x,x^{'})=W^{(1)+}_{33^{'}}(x,x^{'})=W^+(x,x^{'})
\eeq
where $W^+(x,x^{'})$ is the Wightman function for the scalar 
field, Eq.~(\ref{scal}). They are expectation values in the Fulling 
vacuum state, which satisfies $\nabla^aA_a^{(+)}|F>=0$, namely 
\beq
W^{(1)+}_{ij^{'}}(x,x^{'})=<F|A_i(x)A_j(x^{'})|F>\nn
\eeq
The verification of this statement from canonical quantization is 
rather messy, due to an apparent divergence in the integral 
representation of the Rindler Wightman functions. This representation 
also appeared in Ref.\cite{higu92-46-3450}, where the 
Fulling-Davies-Unruh thermal bath was shown to be exactly the 
bremsstrahlung radiation emitted by a uniformly accelerated charge.
The value $\be=2\pi$ is distinguished since then
\beq
W^{(1)\pm}_{ij^{'}}(2\pi|x,x^{'})= g_{ij}W^{\pm}(x,x^{'})\nn
\eeq
where $W^{\pm}(x,x^{'})$ is the scalar Wightman function. This is 
just the Wightman function in the Feynman gauge of Minkowski vacuum 
relative to a boosted tetrad.  
 The Wightman functions obey the Ward identity. In the Feynman
gauge this identity states that
\beq
\nabla^aW^{(1)\pm}_{ab^{'}}+\nabla_{b^{'}}W^{\pm}_{gh}=0
\eeq
where $W^{\pm}_{gh}$ is the Wightman function for the ghost fields
$\eta_1(x)$, $\bar{\eta}_2(x)$.
This is actually equal to the scalar Wightman function because the 
ghosts equations of motion are $\Box\eta_{1,2}(x)=0$.
Though uncoupled to the electromagnetic field, their presence is 
essential in the finite temperature theory\cite{bern74-9-3312}. 
The stress tensor relative to the Minkowski vacuum has 
the perfect fluid form\cite{dowk87-36-3742}
\beq
T^{ab}=\frac{1}{720\pi^2\xi^4}\left[\left(\frac{2\pi}{\be}
\right)^4+10\left(\frac{2\pi}{\be}\right)^2-11\right][4v^av^b+g^{ab}] 
\eeq
%\end{enumerate}
%\end{enumerate}

\section{Discussion}

The previous {\em non} thermal Wightman functions were obtained from canonical 
quantization by calculating explicitely the integral representations 
of the field operators vacuum expectation values. Conversely, 
in the case of finite $\be$, we used different 
methods depending on the value of the 
spin. In fact, for scalar and spinorial fields one can employ 
the method of images, obtaining the corresponding 
non thermal Wightman functions written above. 
One can also implement the canonical formalism, within the Feynmann 
gauge, in the photon case. Then the integral representation of 
$W^{(1)}_{00'}$ and $W^{(1)}_{11'}$ follows from the normal modes 
decomposition of the field operator $A_a$, in the form
\beq
W^{(1)}_{00'}(x,x') = - W^{(1)}_{11'}(x,x') = <F| A_0(x) A_{0'}(x')|F>
= -<F| A_1(x) A_{1'}(x')|F>=\nn
\eeq 
\beq
=\frac{1}{4\pi^{4}} \:
\int_{\R^{2}} dk_{t}\int_{0}^{+\infty} d\omega\:\frac{\sinh \pi\omega}{
k^{2}_{\perp}}\:D\:
 K_{i\omega}(|k_t|\xi) K_{i\omega}(|k_t|\xi')\:
e^{ik_{t}\cdot (x_{t}-x'_{t})}\: e^{-i\omega(\tau-\tau'-i\ep)}\:,\nn
\eeq
where $K_{i\omega}(x)$ is the well-known Mc Donald function of imaginary 
index and the operator $D$ is defined as
\beq
D = \frac{1}{\xi \xi'} [-\partial_{\tau}\partial_{\tau'}+\xi\partial_{\xi}
\xi'\partial_{\xi'}]\:. \nn
\eeq
One can solve the above integral (and the more trivial integral 
representations corresponding to the remaining components) obtaining 
just the non thermal Wightman functions in Eq.s (15), (16), (17). 
Then, the sum over images method  produced the following result 
quite trivially
\begin{eqnarray}
\tilde{W}^{(1)}_{00'}(\be|x,x')&=&-\tilde{W}_{11'}^{(1)}(\be|x,x') \:\:=\:\:  
 W^{(1)}_{00'}(\be|x,x') - \frac{1}{4\pi\be \xi\xi'}\:,\label{anomalous} \\
\tilde{W}^{(1)}_{ab'}(\be|x,x') &=& W^{(1)}_{ab'}(\be|x,x')\:\:\:\:
\mbox{in all the remaining cases}\:, \nn
\end{eqnarray}
where $\tilde{W}^{(1)}_{ab'}(\be|x,x')$ indicates the sum over images result 
arising from $W^{(1)}_{ab'}(x,x')$ and the functions $W^{(1)}_{ab'}(\be|x,x')$ 
were defined by Eq.(10)-(14). Surprisingly then, due to the anomalous 
term in Eq.(\ref{anomalous}), the periodicity sum 
of the zero temperature Wightman functions so obtained 
fails to reduce to the Minkowski Wightmann functions when $\be=2\pi$ and 
thus fails to reproduce exactly the finite temperature result, since the 
thermal properties of the Minkowski vacuum relative to Rindler time 
translations can be established by independent arguments
\cite{haag84-94-219} and even in a model independent and rigorous way
\cite{sewe82-141-201}. We also observe that the sum over images behaves 
badly as $x_{t} \rightarrow \infty$ because $\tilde{W}^{(1)}_{11'}(\beta)$ and 
$\tilde{W}^{(1)}_{00'}(\beta)$ do not vanish there as one might expect
 (in the case of $\beta= 2\pi$ at least). However the correct non thermal 
Wightman functions are reached in the limit $\beta\rightarrow +\infty$.
\\
A closer scrutiny of the situation  
reveals that the responsibility of the failure is due to the sector 
of the photon Fock space containing Rindler states with negative norm. 
In fact, few calculations prove that the anomalous therm $ \Delta_{ab'}(\beta)
=\tilde{W}^{(1)}_{ab'}(\beta)- W^{(1)}_{ab'}(\beta)$ vanishes when
 this acts as a three-distribution on three-smeared solutions of 
vectorial Klein-Gordon equation built up with physical modes only (and 
having compact support on the Rindler Cauchy surfaces, for example)
\footnote{Using a four-smeared formalism, the anomalous term vanishes 
acting as a distribution on conserved currents defined into the {\em open}
Rindler wedge.}.\\
Nevertheless, the result obtained by periodicity sum obeys
both the wave equation and the Ward identities because $ \Delta_{ab'}(\beta)$ 
is a solution of the vectorial Klein Gordon equation having vanishing divergence. 

On the other hand, the Wightman functions for the field strength 
(thus containing no negative norm states), can be obtained by periodicity 
summing over the zero temperature functions 
 $<F|F_{ab}(x)F_{a^{'}b^{'}}(x^{'})|F>$ because the anomalous term
produces a vanishing field strength. We conclude that, dealing with physical 
quantities, the anomalous term as no consequences because it 
represents a gauge ambiguity.\\
Therefore it is possible to drop completely the anomalous
term $\Delta_{ab'}(\beta)$ in the result obtained by summing over 
images, giving just the Wightman functions appearing in Eq.s (10)-(14) 
which reduce to the Minkowski Wightman functions when $\beta=2\pi$.\\
Independently on the method of images, the Wightman functions of Eq.s (10)-(14) 
can be obtained extending Dowker tecnique to handle the scalar Green function 
on a conical space\cite{dowk87-36-3095}. The Rindler metric 
tensor in euclidean time with $\be$ periodicity 
just represents a conical space of the form ${\cal C}_{\be}\times\R^2$, 
${\cal C}_{\be}$ being a two dimensional cone with deficit angle 
$\ga=2\pi-\be$. The Green function of a vector field can then be 
obtained in closed form on the cone after which by analytic 
continuation back to real time one gets precisely Eq.s (10)-(14).\\
It is interesting to observe that, differently from the method of images,
this euclidean approach forces
automatically the Wightman functions to behave correctly at infinity. A 
complete calculation for photons and gravitons was also presented in 
Ref.\cite{alle92-45-4486}, for the case of a cosmic string background 
in which the conical singularity was rounded off. Upon translating
their results to Rindler space, one finds complete agreement. To 
conclude, great care is necessary to deal with gauge fields in 
accelerated frames and in covariant gauges. Related difficulties have 
also been encountered in \cite{kaba95-453-281} and precisely in the 
same context.
  
The thermal stress tensor was obtained by the cited authors using the 
point splitting procedure. Here we give an independent argument 
which is based on the 
old observation\cite{cand79-19-2902} that the manifold ${\cal M}=\R\times
H^3$, 
with the natural product metric, is conformal to Rindler space.
Here $H^3$ is the hyperbolic three space carrying a metric 
with constant negative curvature (an 
extensive discussion of conformally invariant quantum 
field theory in hyperbolic universes has also been given in 
Ref.\cite{bunc78-18-1844,brow86-33-2840}).

The one-loop partition function (per unit volume) for a thermal state in 
${\cal M}$ will be determined by the density of one-particle states in 
$H^3$, denoted $\nu^{(s)}(\om)$ for a spin $s$ field. Indeed 
\beq
\log Z^{(s)}(\be|\xi)=\xi^{-3}\int_0^{\ii}\log\left(1\pm 
e^{\be\om}\right)\nu^{(s)}(\om)d\om -\be U
\label{part}
\eeq
the factor $\xi^{-3}$ coming from the optical space volume element.
$U$ is the vacuum energy density, the only quantity that needs a 
renormalization prescription in this contest. The conformal 
transformation back to Rindler only adds a $\be$-linear 
term\cite{dowk86-33-3150,dowk78-11-895}, 
which may be absorbed into the definition of $U$.
The density of states is thus the crucial quantity. In $H^N$ and for 
the Laplace-Beltrami operator, it has 
long been known by mathematician where it is known as the Harish-Chandra 
or Plancherel measure (see Ref.\cite{camp90-196-1} and Ref.s therein for 
a detailed account). In $H^3$ it is
\beq
\nu^{(0)}(\om)=\frac{\om^2}{2\pi^2}
\eeq
\beq
\nu^{(1/2)}=\frac{(\om^2+1/4)}{2\pi^2}
\eeq
\beq
\nu^{(1)}=\frac{(\om^2+1)}{\pi^2}
\eeq
where the $s=1$ case holds for transverse vector fields in $H^3$ 
(this corresponds to the Coulomb choice of gauge in $\R\times H^3$).
The partition function is now easily computed from Eq.~(\ref{part}). 
We give the details for $s=0$ only, the other cases being similar. 
We obtain
\beq
\log Z(\be|\xi)=\frac{\pi^2}{90\xi^3}\be^{-3}-\be_T U(\La), \hs 
\be_T=\xi\be
\eeq
where $\be_T$ is the Tolman inverse temperature and
\beq
U(\La)=\frac{1}{4\pi^2\xi^4}\int_0^{\La}\om^3d\om\nn
\eeq
ie the regularized vacuum energy density. The linear term will not affects
the entropy density while the energy density must vanishes at $\be=2\pi$, 
since this would correspond to the
scalar vacuum in Minkowski space-time, whose energy density is defined 
to be zero in order to realize the Poincar\'e symmetry.
The zero point of entropy will 
also vanishes at $\be=\ii$ since the Fulling vacuum is a pure states. Once 
the zero point of entropy and energy density have been fixed, there is 
no further room left and all the thermodynamics densities are fixed.
Thus we get the renormalized energy density and pressure
\beq
u(\be)=3p=\frac{1}{480\pi^2\xi^4}\left[\left(\frac{2\pi}{\be}
\right)^4-1\right]
\eeq
the entropy density
\beq
s(\be)=\frac{4\pi^2}{90\xi^3}\be^{-3}
\eeq
and the free energy density
\beq
f(\be)=-\frac{\pi^2}{90\xi^4}[\be^{-4}+3(2\pi)^{-4}]
\eeq
The cut-off dependence is now disappeared. Why should not we define 
the zero of entropy at $\be=2\pi$ which also is a pure state, namely the 
Minkowski vacuum? The reason is a well known consequence of quantum 
theory, first discovered by von Neumann\cite{neum55b}, 
that a subsystem of a 
system in a pure state may has a non zero entropy if only 
the subsystem is being observed. Now while it is true that at 
$\be=2\pi$ we are computing quantities in the Minkowski vacuum, we 
are actually probing only the right hand side Rindler wedge 
since the field operators from which the above results were derived 
were restricted over there. Notice that the zero point free energy is 
equal to the zero point energy and 
that the Gibbs relation $Ts=u+p$ gets modified to 
$Ts=(u-u_0)+(p-p_0)$, in accord with thermodynamic. The total entropy 
is infinite even when $\be=2\pi$, this being the normal behaviour which 
is associated with acceleration horizons. It has been shown that the 
thermodynamic entropy as given above is the same as the entanglement 
entropy\cite{isra76-57-107,kaba94-329-46,sewe82-141-201}, in accord 
with von Neumann ideas.
Eq.~(\ref{tt}) for the stress tensor can now be 
derived by noting that the energy density and pressure must 
be the eigenvalues of the stress tensor in an orthonormal vierbein.

\end{document}